\documentclass[times, 10pt, twocolumn]{article}
\usepackage{latex8}

\newtheorem{definition}{Definition}
\newtheorem{lemma}{Lemma}

\newtheorem{theorem}{Theorem}
\newtheorem{remark}{Remark}
\newenvironment{proof}{\noindent{\bf Proof:  }}
   {\hfill\rule{2mm}{2mm}}

\newcommand{\ket}[1]{|#1\rangle}
\newcommand{\bra}[1]{\langle #1|}
\newcommand{\braket}[2]{\langle #1|#2\rangle}
\newcommand{\Trace}[1]{\rm{Tr\left({#1}\right)}}
\newfont{\scripty}{cmsy10 at 10pt}
\newcommand{\elt}{\in}
\usepackage{amsmath}

\begin{document}
\title{A Quantum Measurement Scenario which Requires Exponential Classical Communication for Simulation}
\author{Allison Coates \\ Computer Science Division \\University of California at Berkeley\\
allisonc@cs.berkeley.edu\\}


\maketitle
\thispagestyle{empty}

   \begin{abstract}

In this paper we consider the following question: how many bits of
classical communication and shared random bits are necessary to
simulate a quantum protocol involving Alice and Bob where they share
$k$ entangled quantum bits and do not communicate at all.  We prove
that $2^k$ classical bits are necessary, even if the classical
protocol is allowed an $\epsilon$ chance of failure.

 \end{abstract}

\Section{Introduction}
\label{intro}
Quantifying the power of shared entanglement is a fundamental goal of
quantum information theory. In this paper, we compare the power of
shared entanglement to the power of classical correlations with
communication by considering the following restricted type of quantum
protocol, which we call {\bf sampling-using-entanglement}: Alice and
Bob are two parties which are not allowed to communicate with each
other, but they share $k$ entangled quantum bits. Each party makes a
generalized measurement on his quantum bits.

We compare this protocol to a classical simulation of it, where Alice
and Bob use classical correlations while communicating classical bits
to sample from a distribution that is $\epsilon$-close of the
distribution resulting from the quantum protocol.

The cost of the classical simulation is the sum of the number of
shared random bits and the number of classical bits communicated
between the two parties.  The number of shared random bits is the
classical analogue to the amount of shared entanglement, while
classical communication serves to augment the limited power of
classical correlations.

Our main result is that $2^k$ classical bits are needed to simulate
the distribution that arises from the quantum protocol's measurement
on $k$ entangled qubits.  Our result is based on the analysis of
sampling protocols for the set-disjointness problem DISJ(x,y)
\cite{Umesh:sampling}. In that paper, the authors demonstrate that the
set-disjointness problem can be sampled using ${\rm O}
\left(\log(n) \log(1/ \epsilon) \right)$ qubits of communication
between two parties, while classical sampling of set-disjointness
requires $\Theta(\sqrt{n})$ classical bits, whether those bits are
communicated on-line during the protocol, or are a cache of prior
correlations.  We cast these results in a different light, showing how
this result holds for a protocol in which the parties share entangled
bits. In particular, we demonstrate an efficient algorithm for
sampling the set-disjointness function DISJ(x,y) using $O\left(\log(n)
\log(1/\epsilon) \right)$ entangled bits and no other communication, which is
again an exponential improvement over the classical sampling
complexity of set-disjointness. These results show that there exist problems
for which classical simulation of an entangled state would require an
exponential amount of communication, demonstrating that classical
correlations are less powerful than quantum correlations.

Another intruiging element of this result is that in this quantum
protocol each party maintains some secret information about their
respective subset, so that parties do not have full information about
the other parties' output. Such a protocol is fundamentally different
than known classical algorithms, where cryptographic schemes are
necessary for $A$ and $B$ to stay ignorant about each other's outputs.

Within the community of quantum communication complexity research,
there is growing body of work addressing the cost of simulation of
entanglement using classical bits\cite{Brassard-Cleve-Tapp:CostofSim},
\cite{DaveCleveOthers:ExpectedSim},
\cite{CP:ClassAnalog}, \cite{Steiner:finitebit},
\cite{CCGM:Classicalteleport}, and similar work in the communication
complexity of entanglement \cite{BCV:CC}. The work of Brassard et.
al.  \cite{Brassard-Cleve-Tapp:CostofSim} showed that 8 classical bits
of communication suffice to exactly simulate a Von Neumann measurement
on a Bell pair if an infinite random string is shared between two
parties.  That paper also shows a partial function for which an exact
classical simulation of an entangled system of $n$ qubits would
require $2^n$ classical bits of communication.  However, they left
open the question of whether any classical simulation existed on a
more robust quantum measurement scenario when that simulation allowed
small error, or was limited in the amount of bits available in prior
correlation. Massar et. al.
\cite{DaveCleveOthers:ExpectedSim} showed that no protocol restricted
to a finite amount of prior correlation could simulate the Bell pair, but
that Bell states can be simulated without an infinite random string if
the expected classical communication is finite.


Our work differs from the models studied above because we consider the
the communication complexity of sampling rather than of functions or
partial functions.  Further, we demonstrate a protocol that
is robust to small errors.  Lastly, as shown in Massar et.
al.\cite{DaveCleveOthers:ExpectedSim} an infinite amount of shared
randomness is too powerful to be used as a resource.  As such, we
quantify the total amount of classical information that is used in the
protocol, counting the amount of prior correlations as well as the
amount of communicated bits. Finally, we believe that the sampling
model is more appropriate for investigation of quantum communication
complexity problems since measurement of a quantum mechanical system
is inherently a sampling process.

In \ref{prelim} we discuss classical and quantum sampling protocols and
introduce our restricted quantum protocol for
sampling-using-entanglement.  In \ref{res} we state the problem of 
set-disjointness, and introduce an sampling-with-entanglement protocol
for it.
Section \ref{proof} contains the main theorem proof of the
upper bound on the communication complexity of
sampling-using-entanglement for the set-disjointness problem,
while \ref{lemma} presents the proof of the main lemma for this result.  In
\ref{classical} we present the classical sampling results for
set-disjointness, demonstrating an exponential gap between the
communication complexity of sampling-using-entanglement and
communication complexity of classical sampling.

\Section{Preliminaries}\label{prelim}
Consider a scenario in which two parties, Alice and Bob wish to output
samples according to some known joint probability distribution.  They
share ${\ket{\psi_{AB}}}$, a bipartite entangled state of $k$ quantum bits,
but they cannot communicate classically with each other. A simple
protocol, labelled $\left\{U_A, U_B, \ket{\psi_{AB}}\right\},$ samples a
probability distribution as follows: each party performs a unitary
operation, $U_A$ and $U_B$, respectively, on their portions of the
entangled system $\ket{\psi_{AB}},$ and additional ancilla. Then, each
party performs a measurement in the computational basis and outputs
the results.  Since a measurement of a quantum system results in a
classical probability distribution, this protocol samples from the
classical probability distribution ${\scripty \Pi}\left\{ U_A, U_B,
    \ket{\psi_{AB}}\right\}$ induced by performing a measurement $U_A
\otimes U_B$ on $\ket{\psi_{AB}}$. For shorthand, we will refer to a
unitary operation and subsequent measurement in the computational
basis as a {\bf generalized measurement}, and say that the protocol
samples a state $\left(U_A \otimes U_B \right) \ket{\psi_{AB}}$ to mean
that the protocol outputs samples according to the classical
probability distribution ${\scripty \Pi}\left\{ U_A, U_B,
    \ket{\psi_{AB}}\right\}.$

Now imagine Alice and Bob collude; that is, they claim to share an
entangled state, but in reality they have only classical communication
and prior classical correlations available to them. They will again be
given unitary operators and asked to output measurements according to
${\scripty \Pi}\left\{ U_A, U_B, \ket{\psi_{AB}}\right\}$. In order for
them to succeed in convincing a third party that they share
entanglement, they must ``simulate'' this quantum protocol by
outputting results according to a probability distribution that is
$\epsilon$-close to the probability distribution ${ \scripty
\Pi}\left\{ U_A, U_B,\ket{\psi_{AB}}\right\}$.

We now define formally a two-party quantum protocol for the
sampling-using-entanglement of $\ket{\psi_{AB}}$ and the subsequent
classical simulation of such a protocol. To do so, we include an
intermediate definition of q-sampling \cite{Umesh:sampling} where the protocol
that samples $\ket{\psi_{AB}}$ not only outputs values according to the
distribution of $\ket{\psi_{AB}}$ but creates a pure state that is close to
the ${\ket{\psi_{AB}}}$ desired.



\begin{definition}
    Let $\ket{\psi_{AB}}$ be a bipartite entangled state, and let
    $U_A$ and $U_B$ be unitary operators.  A two-party {\bf
    sampling-using-entanglement protocol} $Q$ is a quantum protocol
    executed by Alice and Bob as follows: Alice and Bob perform their
    respective unitary operators on their portions of the
    $\ket{\psi_{AB}}$ and some additional ancilla.  Then, each party
    performs a generalized measurement in the computational basis. The
    two parties do not communicate in any other way.  This protocol
    outputs samples according to the classical probability
    distribution $\scripty{\Pi}\left\{U_A, U_B,
    \ket{\psi_{AB}}\right\}$, defined as probability distribution
    induced by performing a measurement $U_A \otimes U_B$ on
    $\ket{\psi_{AB}}$. Further, we say that $Q$ {\bf q-samples}
    a state $\ket{\Psi}$ if $\bra{\Psi}{(U_A \otimes
    U_B)\ket{\psi_{AB}}} > 1 - \epsilon$. The communication complexity
    of sampling-using-entanglement is the number of quantum bits
    shared between the two parties.
\end{definition}

\begin{definition}
Let $Q$ be a sampling-using-entanglement protocol as above that samples
the distribution $\scripty{\Pi}\left\{U_A, U_B,
\ket{\psi_{AB}}\right\}$.  Let $P$ be a two-party classical protocol
$P$ where each party is computationally unbounded. We say
$P$ is a {\bf classical simulation} of $Q$ if $P$ outputs samples $x$
and $y$ according to a distribution $\scripty
\Pi'(P)$, such that that $| \scripty{\Pi'(P)} - \scripty{\Pi} \left\{ U_A, U_B, \ket{\psi}\right\}|_{TVD} <
\epsilon$, where the norm is the Total Variational Distance.  The
communication complexity of $P$ is $k+m$, the sum of the number of
classical bits of communication $(k)$ and the number of classical bits
of correlation $(m)$.
\end{definition}

The classical simulation does not need to simulate the specific
quantum state or its overall phase information; it only matches the
overall probability distribution.  However, the quantum sampling
protocol we describe is more restrictive, and matches the appropriate
quantum state, as well as the resulting distribution.

In previous literature \cite{Umesh:sampling}, the communication
complexity of classical sampling is defined as the number of classical
bits of communication needed to sample a function.  This definition is
equivalent to the number of bits of correlation needed to sample $f$.


\Section{Discussion}
The main result of this paper is to answer the question of whether all
quantum measurement scenarios can be simulated with only a polynomial
number of classical bits.  We demonstrate that the answer is no, by
showing that already established results in quantum communication
complexity can be applied directly to this problem and provide the answer.

The main technical theorems of this paper were first proved by
Ambainis et. al. \cite{Umesh:sampling} in a paper comparing the
classical and quantum communication complexities of sampling. While we
provide the details of those theorems again for completeness, our main
result is to show that those results apply directly to a new
communication protocol of sampling-using-entanglement. A
sampling-using-entanglement protocol is equivalent to a quantum
measurement scenario: both protocols involve each party performing a
generalized measurement on their portion of the entangled state.

Prior work in quantum information theory has failed to answer the
question of whether all larger non-separable entangled systems could
be simulated with even a polynomial number of bits.  However, by
applying the communication complexity results for sampling, we find
that if such a scenario could be simulated with less than an
exponential number of bits relative to the quantum scheme, then we
could beat the classical sampling bound proven in Ambainis et. al.
Therefore, any classical simulation of such a quantum measurement
scenario must match the classical communication lower bound on
sampling. We use the results of Ambainis et. al. to demonstrate a
quantum measurement scenario for the set-disjointness problem on $n$
entangled qubits cannot be simulated with less than $2^n$ classical
bits of communication or correlation.


\Section{Results for sampling of set-disjointness}\label{res}

We now show that there exists a sampling-using-entanglement quantum
protocol for the set-disjointness problem that uses only
$O(\log{n}\log{1/\epsilon})$ qubits.  This result provides an
exponential speedup over the classical sampling result for
set-disjointness, where $\Theta(\sqrt{n})$ classical bits of
communication and shared randomness are required.  The treatment
provided here is based on the treatment in Ambainis et. al, but
differs greatly in its organization and presentation.

 Formally, the problem of sampling set-disjointness (for size
$\sqrt{n}$) is defined as follows: Given a set $\Omega = \{ 1, \dots,
n\} $, two parties $A$ and $B$ sample the set-disjointness function if
they output uniformly random disjoint subsets $S$ and $T$
respectively, $S, T \in \Omega$, $|S| = |T| =\Theta(\sqrt{n})$.  Let
$\ket{\chi}$ be the superposition representing all possible disjoints
subsets of size $\sqrt{n}$: $\ket{\chi} = \sum_{i,j \cap \emptyset}
\ket{i}\ket{j}$.  A protocol can quantum sample $\ket{\chi}$ if it
creates a state $\ket{\phi}$ which is close to $\ket{\chi}$.  The
trivial sampling-using-entanglement protocol quantum samples this
function: it simply measures $\ket{\chi}$ in the computational basis.
However, the number of qubits necessary to express $\ket{\chi}$ is
$\log{ n\choose \sqrt{n}} \approx \sqrt{n} \log{n}$, which does not
improve upon the classical sampling complexity of this
problem. Instead, we will show that there exists a smaller bipartite
system that contains nearly all of the necessary information about
$\ket{\chi}$.  Specifically, we will construct a pure state
$\ket{\psi}$ of only ${\rm O}(\log(n)\log(1/\epsilon))$ entangled
qubits which when operated upon by local unitary operations $U_A$ and $U_B$
 yields a state which is an $\epsilon$-approximation for
$\ket{\chi}$.

\begin{theorem} 
    There exists a basis $V$ for $\ket{\chi}$, the superposition
    ranging over all possible disjoint subsets of size $\sqrt{n}$ with
    the following property: L
et $\ket{\psi}$ be the projection of
    $\ket{\chi}$ onto the subspace spanned by the largest
    $n^{\log{1/\epsilon}}$ eigenvectors of $V$. Then $\bra{\chi}\left(V
    \otimes V\right)\ket{\psi} > 1 - \epsilon$.
\end{theorem}


\Section{Proof of Main Theorem}\label{proof}
\begin{proof}
We begin by using a matrix representation for states and the unitary operators. 
Consider $\ket{\chi}$ as above, written as 
\[
\ket{\chi} =\sum_{i,j} m_{ij}\ket{i}\ket{j} \hbox{ where } m_{ij} =
\left\{ \begin{array} { r@{\quad:\quad}l} 1 & i \cap j = \emptyset,\\
0 & i \cap j \neq \emptyset \end{array} \right. \] Let $M_{\chi}$ be
the linear operator corresponding to the matrix representation for
$\ket{\chi}$: $M_{\chi} = \left[ m_{ij} \right]$, where $M_{\chi}$'s
rows are indexed by $\ket{i}$ and columns by $\ket{j}$.


\begin{remark}
    {\rm Given a state }$ \ket{\phi} = \sum_{i,j} m_{i,j}
    \ket{i}\ket{j}$, {\rm a matrix representation }$ M_{\phi} = [m_{i,j}]$, {\rm and 
      unitary operators} $U_A$ {\rm and} $U_B$, {\rm the matrix representation
    for the state} $\left(U_A \otimes U_B \right) \ket{\phi}$ {\rm is } $U_A
    M_{\phi} U_B^\dagger$. \footnote{To see the proof of this remark,
    please refer to the Appendix.}
\end{remark}
Using the representation, a rotation of a state $\ket{\psi}$ into the
$V$ basis can be written as $V M_{\psi} V^{\dagger}$.  This notation
is particularly useful because it allows us to compare the norms of
pure states by comparing the trace norms of their matrix
representations. We define the trace norm of a matrix $M$ as $|| M || =
\sqrt{\Trace{M^{\dagger}M}}$.

\begin{remark}
{\rm    Given states $\ket{\psi}$ and $\ket{\chi}$, their corresponding
    matrix representations $M_{\psi}$ and $M_{\chi}$, and a basis $V$,
    then for every $\epsilon$, if ${||V M_{\psi} V^{\dagger} -
    M_{\chi}||}^2 < 2\epsilon$, then $\bra{\chi} \left(V \otimes V
    \right) \ket{\psi} > 1- \epsilon$. }
\footnote{To see the proof of this remark, please refer to the Appendix.}
\end{remark}

With the above two facts in place, it remains to prove that there
exists an appropriately-sized basis for the matrices representing our
bipartite quantum states.

\begin{lemma} There exists a basis $V = \{v_1, \ldots v_n\}$ for 
$M_{\chi}$ such that the linear operator $M_{\psi}$, the projection of
$M_{\chi}$ onto the subspace spanned by $n^{\log(1/\epsilon)}$
eigenvectors of $V$ corresponding to the largest
$n^{\log(1/\epsilon)}$ eigenvalues, is within $2\epsilon$ of
$M_{\chi}$: $${||V M_{\psi} V^{\dagger} - M_{\chi}||}^2 < 2 \epsilon$$
\end{lemma}

The proof then follows.
\end{proof}

Intuitively, this proof follows from the fact that we need only a
small number of basis vectors to closely approximate any vector in the
vector space of the linear operator $M_{\chi}$. We can then rotate
that system of basis vectors back into a larger space and achieve a 
linear operator that is close to $M_{\chi}$.  In subsequent sections
we prove the associated lemma.  


\Section{Classical Bound comparison }\label{classical}

We now return to the issue of simulation of quantum measurement.
Since the sampling-using-entanglement protocol is equivalent to a
quantum measurement scenario that samples set-disjointness, any
classical simulation of this quantum measurement scenario would need
to classically sample set-disjointness. But by Ambainis et. al.
\cite{Umesh:sampling}, the lower bound for classical sampling of
set-disjointness is $\Omega(\sqrt{n})$. Therefore, any classical
protocol simulating the the quantum measurement scenario of
set-disjointness must use $\Omega(\sqrt{n})$ bits of communication.
This demonstrates an exponential gap between
sampling-using-entanglement and classical sampling, and shows that the
result of Brassard et. al. \cite{Brassard-Cleve-Tapp:CostofSim} that a
Von Neumann measurement on a Bell pair could be simulated with a
constant number of classical bits of communication does not scale to
larger quantum measurement scenarios on systems of qubits.

\Section{A cryptographic protocol}\label{crypto}
This protocol shows that given a small initially entangled state, by
performing unitary operations on that state and some ancilla, we can
rotate into a larger space. But because the original entangled state
resided on a smaller subspace, parties $A$ and $B$ do not have full
knowledge of each other's final output. Indeed, the are only aware of
$O(\log(n)\log{(1/\epsilon)})$ bits of information of each others
output, rather than $\sqrt{n}$ bits.  This is a fundamental difference
over classical information theory, where the existence of
cryptographic one-way functions is necessary to achieve the same
``hidden'' bits in a sampling protocol.

We believe that this amount of hidden information could be used to
create future cryptographic protocols for property testing. Those
result will be shown in a later version of this paper.  We also
believe that this imbalance between hidden information and classical
simulation results is an interesting area for future research into the
power of entanglement. This idea addresses some of the concerns
expressed in Collins and Popescu \cite{CP:ClassAnalog}, who raised the question of what
particular behavior of entanglement is quantum-mechanical.  We
believe that the hidden bits here may directly answer this question.


\Section{Proof of Lemma 1}\label{lemma}

\begin{proof}
We begin by assuming that such a basis exists,
and consider the $t$ eigenvectors that correspond to the largest 
$t$ eigenvalues. We derive conditions on the value of $t$ 
that succeeds in achieving an approximation bound of $2 \epsilon$.

Let $V$ be a basis of eigenvectors for $M_{\chi}$: $M_{\chi} = V
\Lambda V^{\dagger}$, where $\Lambda$ is the matrix of eigenvalues.  
Let $V_{t}$ denote the subspace defined by the first $t$ eigenvectors of $V$.
We define $M'$  as the linear operator created by projecting $M$ onto the subspace of the
first $k$ eigenvectors in $V$.

        \begin{eqnarray*}
            {\|| V M' V^{\dagger} - M_{\chi} \|} ^2 
            &=& {\|| V M' V^{\dagger} - V \Lambda V^{\dagger}\|}^2 \\
            &=& {\| V (M' - \Lambda ) V^{\dagger}\|}^2\\
            &=& {\| M' - \Lambda \|}^2 \\
            &=& {\| V_{t} M_{\chi} V_{t}^{\dagger} - \Lambda \| }^2  \\
            &=& \sum_{i>{t}}^{|V|} {\lambda_i}^2
        \end{eqnarray*}

Therefore, $|| V M' V^{\dagger} - M_{\chi} || ^2 < 2 \epsilon$ if
$\sum_{i>{t}}^{|V|} {\lambda_i}^2 < 2\epsilon$.  We continue by
characterizing the eigenvalues of $M_{\chi}$, and use this information
to derive a bound on $t$ to ensure $\sum_{i>{t}}^{|V|} {\lambda_i}^2 <
2\epsilon$.

$M_{\chi}$ can be written as a linear combination of the ones matrix
and a related matrix $B=[b_{ij}]$ of size $N = {n \choose k}$ by
$N={n\choose k}$ , whose rows are indexed by $\ket{i}$, columns by
$\ket{j}$, and $b_{ij} = 1 \mbox{ if }i \cap j = \emptyset, -1 \mbox{
if } i \cap j \neq \emptyset$. Since $M_{\chi}$ is a linear combination
of $B$ and the ones matrix, $B$ has the same eigenvectors as $M_{\chi}$.


$B$ is real and symmetric, and therefore has an orthonormal set of
 eigenvectors $W= \{w_1, \ldots, w_{N-1}\}$.  Using a result of
 Lovasz for $B$ \cite{Lovasz:ShannonCapacity}, we characterize
the eigenspaces of $B$ as follows:
        
        \begin{lemma} \cite{Lovasz:ShannonCapacity} $B$ has $k+1$
            eigenspaces. Eigenspace $E_0$ is of dimension $1$ and
            contains the all $1's$ vector. $E_i$ has dimension ${n
            \choose i} - {n \choose {i-1}}$. The typical eigenvector
            in $E_i$ is indexed by $x_1 , \ldots , x_{2i-1} , x_{2i}
            \in \{1,\dots,n\}$.  The corresponding eigenvector $e$
            (unnormalized) is given by: $e_S = 0$ if there is an index
            $j: |S \cap \{x_{2j-1}, x_{2j}\}| \neq 1$, otherwise $e_S
            = \Pi_j (-1)^{|S \cap x_{2j}\}|}$.  The corresponding
            eigenvalues are $\lambda_0 = \frac{2 {{n-k} \choose k} -
            {n \choose k}}{{n \choose k}}$, and $\lambda_i = \frac{2
            {{n-k-i} \choose {k-i}} }{{n \choose k}}$.  \end{lemma}
      
      The first $g$ eigenspaces contain less than $g {{n}\choose{g}}
      \leq n^{g+1}$ eigenvectors. Setting $t = n^{g+1}$, we wish to
      show that we can pick $g$ small enough to satisfy $\sum_{i>
        t}^{|V|} {\lambda_{i}^2} < 2\epsilon$.
       
A projection of a row of $B$ onto subspace $E_i$ is the sum of the
row of $B$ onto each vector $w$ spanning $E_i$.  Let $q_i $
be the square of the length of the projection of a row $b$ onto the
subspace $E_i$. Since the value of such a projection is
$\lambda_i$, then ${q_i} = \sum_{w \elt E_i}{\lambda_i}^2$. 
Therefore, $$\sum_{i>t}^{|V|} \lambda_{i}^2 \leq \sum_{i > g}^{k+1} \sum_{w \elt E_i} \lambda_i^2
= \sum_{i > g}^{k} q_i.$$
Since each $E_i$ has dimension ${n \choose i} - {n \choose {i-1}}$, $E_i$ has ${n \choose i} - {n \choose {i-1}}$ eigenvectors, each with value
$\lambda_i = \frac{2 {{n-k-i} \choose {k-i}} }{{n \choose k}}$.
Therefore, $$q_i = \left({n \choose i} - {n \choose {i-1}}\right){\left(\frac{2 {{n-k-i} \choose {k-i}} }{{n \choose k}}\right)}^2.$$

Next, we show that the projections decay rapidly, by showing the relative change between $q_i$ and $q_{i+1}$.
            \begin{eqnarray*}
                \frac{q_{i+1}}{q_i} 
                &=& \frac{ \left({n \choose \left(i+1\right)} - {n \choose {i1}}\right)\left(\frac{2 {{n-k-\left(i+1\right)} \choose {k-\left(i+1\right)}} }{{n \choose k}}\right)^2}
                { \left({n \choose i} - {n \choose {i-1}}\right)\left(\frac{2 {{n-k-i} \choose {k-i}} }{{n \choose k}}\right)^2}\\
                &\leq&\frac{2n}{i+1} \cdot \frac{\left(k-i\right)^2}{\left(n - k - i\right)^2} \\
                &\leq&\frac{2n}{i+1} \cdot \left(\frac{2k}{n}\right)^2
\end{eqnarray*}
If $k = \Theta(\sqrt{n})$,
then $\frac{q_{i + 1}}{q_i} = O(\frac{1}{i+1})$.  Therefore, $q_g \leq \frac{c^g}{g}$. 

Choosing $g$ to be  $k\log{n}\log(1/\epsilon)/\log\log(1/\epsilon))$,
$$\sum_{i > g}^{k} q_i= \sum_{i > g}^{k} c^i/i < c^g/g! < \epsilon.$$ 
Since $\sum_{i>t}^{|V|} \lambda_{i}^2 \leq \sum_{i > g}^{k+1} q_i$,
$$\sum_{i>t}^{|V|} \lambda_{i}^2 \leq \epsilon.$$
With $g = O(\log(1/\epsilon)/\log\log(1/\epsilon))$, the total number
of eigenvectors needed is bounded above by $n^{g+1} = O(n^{\log
1/\epsilon})$, and therefore only $O( \log n \log 1/\epsilon)$
entangled qubits are needed to perform the protocol. This completes
the proof.
\end{proof}
\Section{Acknowledgements}
The author wishes to thank Umesh Vazirani, Richard Josza, and Jennifer
Sokol for numerous helpful discussions. This research is supported by
a Siebel Scholars Fellowship and DARPA Grant F30602-00-2-060.

\bibliography{quantum}

\appendix
\Section{Proof of Remark 1}

We prove 
that given a bipartite state $\ket{\psi}$ and local unitary operations
$U$ and $V$, we can write the result of $\left(U \otimes V\right)
\ket{\psi}$ as $U M V^{\dagger}$ where $M$ is the matrix representing
$\ket{\psi}$.

\begin{proof}

A and B want to perform $U \otimes V$ on a bipartite state $\ket{\phi} = \sum_{ij} \phi_{ij} \ket{i}\ket{j}$.
The coefficient $\phi_{ij}$ refers to the probability amplitude that a measurement of the state $\ket{\phi}$  will yield $\ket{i}$ to party $A$ and $\ket{j}$ to party $B$.
Performing the unitary operation $U \otimes V$ on $\ket{\phi}$ 
changes the coefficients to $\ket{\phi}'$ according to the rule
\begin{eqnarray*}
\phi_{ij}' &=&\braket{(i  \hbox{th row of } U \otimes j\hbox{th row of }V}{\phi}\\
&=& \sum_k \sum_m U_{ik}V_{jm} \phi_{km}\\
&=& \sum_k U_{ik} \sum_m V_{jm} \phi_{km} \\
&=& \sum_k U_{ik} \sum_m \phi_{km} V_{jm}
\end{eqnarray*}

Now, we show that in the matrix formulation,
$U \otimes V\ket{\phi} = U M V^\dagger$.  Beginning with a matrix $M$,
we show the values  of the matrix $UMV^{\dagger}$.
\begin{eqnarray*}
(UM)_{ij} &=& \sum_{k=1}^n U_{ik} M_{kj}\\
(AV^T)_{ij} &=& \sum_{m=1}^n A_{im} V_{jm}\\
(UAV^T)_{ij} &=& \sum_k U_{ik} \left(\sum_m A_{km} V_{jm}\right)\\
(UMV^T)_{ij} &=& \sum_k U_{ik} \left(\sum_m M_{km} V_{jm}\right)
\end{eqnarray*}

Hence, these two formulations are equal.
\end{proof}

\Section{Proof of Remark 2}


\begin{proof}

Note that $\Trace{M_{\chi}^{\dagger}M_{\chi}} = \braket{\chi}{\chi}$,
since $\Trace{M_{\chi}}^{\dagger}\,M_{\chi} = \sum_{ij}
{\left(M_{\chi}\right)}_{ij}.$ But since this matrix consists of all
the elements of ${\ket{\chi}}$, this result is $\sum_{ij}
\left(M_{\chi_{ij}}\right) = \braket{\chi}{\chi}.$

We rewrite  ${||V M_{\psi} V^{\dagger} - M_{\chi}||}^2 $ in terms of
 the trace norm: 
\begin{eqnarray*}
\lefteqn{{||V M_{\psi}V^{\dagger} - M_{\chi}||}^2 = }\\
&& \Trace{{\left(VM_{\psi}V^{\dagger}\right)}^{\dagger}{\left(VM_{\psi}V^{\dagger}\right)}} - \\  
&& 2\Trace{M_{\chi}^{\dagger} V M_{\psi} V^{\dagger}} + \Trace{{M_{\chi}}^{\dagger}M_{\chi}}
\end{eqnarray*} Since the trace is invariant
 under basis transformations, and $V^{\dagger}V = I$, 
\begin{eqnarray*}
\Trace{\left(V  M_{\psi}V^{\dagger}\right)}^{\dagger}{\left(VM_{\psi}V^{\dagger}\right)} 
&=& \Trace{ \left(V M_{\psi}^{\dagger} V^{\dagger} V M_{\psi}  V^{\dagger} \right) }\\
&=& \Trace{ \left( M_{\psi}^{\dagger} M_{\psi}  \right) } \\
&=& \sum_{ij} \left(M_{\psi_{ij}}\right) =  \braket{\psi}{\psi}
\end{eqnarray*}

Continuing, we see that 
\begin{eqnarray*}
- 2 \Trace{ V M_{\psi}^{\dagger} V^{\dagger}  M_{\chi}} 
&=& - 2 \Trace{ V M_{\psi}^{\dagger} V^{\dagger} V \Lambda  V^{\dagger} }\\
&=& - 2 \Trace{ M_{\psi}^{\dagger} \Lambda }\\
&=& - 2  \bra{\psi, 000}\ket{\chi}
\end{eqnarray*} 
where $\bra{000}$ represents an
 ancilla, so that the $\ket{\chi}$ and $\ket{\psi, 000}$ lie in the
 same vector space. 

Therefore, 
\begin{eqnarray*}
{||V M_{\psi} V^{\dagger} - M_{\chi}||}^2 &=& \braket{\psi}{\psi} + \braket{ \chi}{\chi} - 2\braket{\psi, 000}{\chi} \\
&=& 2 - 2\braket{\psi,\, 000}{\chi} \\
&<& 2\epsilon
\end{eqnarray*}

Therefore, $\braket{\chi}{\psi,\,000}> 1 - \epsilon$. Since unitary
operations on pure states cannot change their length,
$\braket{\chi}{\psi,\, 000} = \bra{\chi}{V \otimes V}\ket{\psi,\, 000} > 1- \epsilon$.
\end{proof}

\end{document}